\def \Rmip {R_\textrm{mip}}
\def \rhoSM {\rho_\textrm{SM}}
\def \etaSM {\eta_\textrm{SM}}
\def \Emax {E_{max}}

\documentclass[epj]{svjour}
\usepackage{graphics}
\usepackage{epsfig}
\usepackage{amsmath}
\begin{document}
\title{Results from the ICARUS T600 module}
\subtitle{A measurement of the $\mu$ decay spectrum}
\author{Javier Rico for the ICARUS Collaboration
\thanks{L'Aquila,  LNF, LNGS, Milan, Naples, Padova, Pavia, Pisa, 
Torino, ETH Z\"urich, Beijing, Katowice, Krakow, Warsow, Wroclaw,
UCLA, Granada, INR Moscow} } 
\offprints{Javier.Rico@cern.ch}          
\institute{Institut f\"ur Teilchenphysik\\
ETH H\"onggerberg, Z\"urich, Switzerland}
\date{Received: date / Revised version: date}
%
\abstract{
We have studied the $\mu$ decay energy spectrum from a sample of
stopping $\mu$ events acquired during the test run of the ICARUS T600
prototype. This detector allows the spatial reconstruction of the
events with fine granularity, hence the precise measurement of the
$\mu$ range and $dE/dx$ with high sampling rate. This information is
used to compute the correction factors needed for the calorimetric
reconstruction. The Michel $\rho$ parameter is then measured by
comparison of the experimental and Monte Carlo simulated $\mu$ decay
spectra, obtaining $\rho = 0.72\pm 0.06 \textrm{ (stat.)} \pm 0.08
\textrm{ (syst.)}$.
} 
\maketitle

\section{Introduction}

The sample of events in which a muon enters the detector, stops and
eventually decays in the detector's sensitive volume --hereafter
called \emph{stopping muon} sample-- constitutes an important
benchmark to evaluate the physics performance of the ICARUS
detector. Because of their simple geometry, stopping muon events are
relatively easy to reconstruct in space, which allows the computation
of the different correction factors needed for the calorimetric
reconstruction. Thus, we can study the muon decay spectrum and measure
the Michel $\rho$ parameter, which constitutes the first physics
measurement performed with the ICARUS novel detection technology, and
proves that the technique is mature enough to produce competitive
physics results.

Muon decay was first described in a model-independent way by
Michel~\cite{MICHEL}, using the most general local, derivative-free,
lepton-number conserving, four fermion interaction. For unpolarized
muons, the decay probability is given by:
\begin{eqnarray}
\frac{dP}{dx}(x;\rho,\eta) & = & \frac{1}{N} x^2 \left(3 (1-x) +
\frac{2}{3}\rho(4x-3)+ \right. \\ \nonumber
& & \left. 3\eta\frac{m_e}{\Emax} \frac{1-x}{x} +
\frac{1}{2} f(x) + \mathcal{O}(\frac{m_e^2}{\Emax^2})\right) 
\label{eq:michel}
\end{eqnarray}
$x=\frac{E_e}{\Emax}$ is the \emph{reduced} energy (ranging
from $m_e/\Emax$ to 1); $E_e$ and $m_e$ are respectively the total
energy and mass of the electron produced in the decay;
$\Emax=52.8$~MeV is the end-point of the spectrum; $f(x)$ is the
term accounting for the first order radiative corrections assuming a
local V-A interaction~\cite{RADCORR}; finally, $\rho$ and $\eta$ are
the so-called Michel parameters, defined in terms of bilinear
combinations of the coupling constants of the general four fermion
interaction, and hence depending on the type of interaction governing
the decay process. For the Standard Model (SM) V-A interaction, the
parameters take the values $\rhoSM = 0.75$ and $\etaSM = 0$.

\section{The ICARUS T600 detector}

\begin{figure}[!tp]
\begin{center}
\epsfig{file=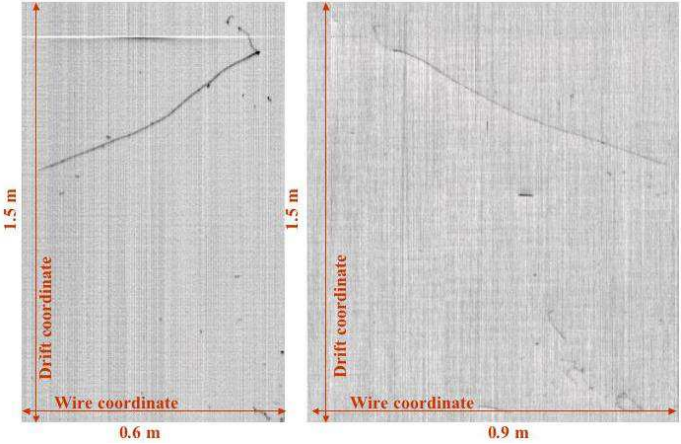,height=6cm}
\end{center}
\caption{Run 966 Event 8 Right chamber: muon decay event views
corresponding to the Collection (left) and second Induction (right)
wire planes.}
\label{fig:stopmu}
\end{figure}

ICARUS T600~\cite{T600} is a large cryostat divided in two identical,
adjacent half-modules of internal dimensions $3.6 \times 3.9 \times
19.9$ m$^3$, each containing more than 300~t of liquid argon
(LAr). Each half-module houses an internal detector composed by two
Time Projection Chambers (TPC), the field shaping system, monitors,
probes, PMT's, and is externally surrounded by a set of thermal
insulations layers. Each TPC is formed by three parallel planes of
wires, 3~mm apart, oriented at $0, \pm60^\circ$ angles, of 3~mm pitch
parallel wires, positioned onto the longest walls of the
half-module. The cathode plane is parallel and equidistant to the wire
planes at each TPC. A high voltage system produces a uniform electric
field, perpendicular to the wire planes, forcing the drift of the
ionization electrons (the maximum drift path is 1.5~m).

The ionization electrons produced in the LAr active volume drift
perpendicularly to the wire planes due to the applied electric field,
inducing a signal ({\it hit}) on the wires near which they are
drifting while approaching the different wire planes. By appropriate
biasing, the first set of planes can be made non-destructive ({\it
Induction} planes), so that the charge is finally collected in the
last plane ({\it Collection} plane).  Each wire plane provides a
two-dimensional projection ({\it view}) of the event, where the
position in one coordinate is constrained by the hit wire, while the
signal timing with respect to the trigger gives the position along the
drift direction.

\section{Data selection and reconstruction}

\label{sec:recon}
\label{sec:calrec}

Data were acquired during the T600 technical run~\cite{T600}. The
off-line data selection was carried out by visual scanning using
topological criteria. A total of 5830 triggers were scanned,
containing 4548 stopping muon events, out of which 3370 were further
selected by the preliminary quality cuts.

The spatial reconstruction of the muon tracks is needed in order to
compute the corrections entering the calorimetric reconstruction of
the events. A detailed description of the spatial reconstruction tools
has been reported elsewhere~\cite{T600,THESIS}. 

The ionization charge is precisely measured in the Collection wire
plane. The energy associated to a given hit is related to the
collected charge by means of 
\begin{equation}
E = \frac{CW}{R}\ e^{(t-t_0)/\tau_e}\, Q
\end{equation}
where $C$ is the calibration factor~\cite{CALIBRATION};
$W=23.6^{+0.5}_{-0.3}\textrm{ eV}$ is the average energy needed for
the creation of an elec\-tron-ion pair~\cite{MIYAJIMA}; $R$ the
electron-ion recombination factor; ($t-t_0$) the time of drift of the
electrons; $\tau_e$ the drift electron lifetime, which parametrizes
the attachment of drift electrons to impurities in LAr; and $Q$ the
measured charge. $R$, $t_0$ and $\tau_e$ are extracted from the
reconstructed muon tracks~\cite{THESIS,STOPMU}, essentially by tuning
them so that the measured energy corresponds to the theoretical
expectation for stopping muons. This method determines the electron
energy in a bias-free way, since all the correction parameters are
tuned using exclusively muon tracks. For $\tau_e$ we have obtained
values ranging from 1.2 to 1.7~ms (depending on the data taking
period), meaning free paths for drift electrons between 1.9 and 2.6~m,
to be compared with the maximal drift distance 1.5~m. The measured
recombination factor is $\Rmip = 0.640
\pm 0.013$ for minimum ionizing particles and $R^{-1}$ linearly
increasing with slope $0.11 \pm 0.01$~cm/MeV.

\begin{figure}[!tp]
\begin{center}
\epsfig{file=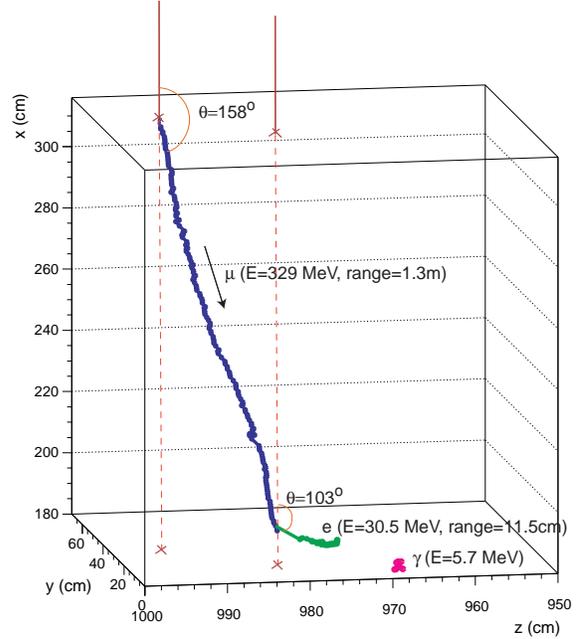,width=7.5cm}
\end{center}
\caption{Run 966 Event 8 Right chamber: fully reconstructed  muon
decay event.}
\label{fig:stopmu3d}
\end{figure}

Figure~\ref{fig:stopmu} shows the initial two-dimensional projections
of a typical stopping muon event, produced by the Collection and the
second Induction planes, respectively. On Figure~\ref{fig:stopmu3d},
the fully reconstructed event is shown.

\section{Results}

\begin{figure}[!tp] 
\begin{center}
\epsfig{file=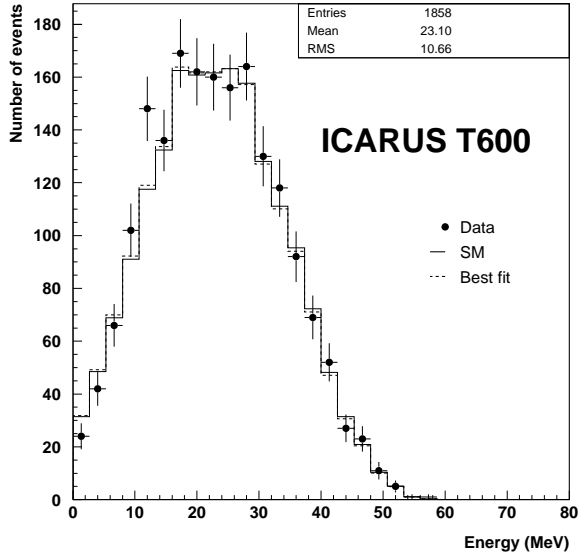,width=8cm}
\end{center} 
\caption{Muon decay spectra in the ICARUS T600 detector. The
plot shows the measured distribution (dots) compared both to the SM
expectation (solid line) and the best fit with $\rho$ and $\eta$ as
free parameters (dashed line).}
\label{fig:eletrackfit} 
\end{figure}

A total of 1858 electrons are available for the determination of the
Michel $\rho$ parameter. The measured spectrum corresponds to the
fraction of the electron energy lost by ionization in LAr since no
attempt to recover the energy loss by bremsstrahlung radiation has
been carried out. We compare the spectrum with the one obtained for a
Monte Carlo (MC) simulated event sample, containing 10000 electron
events from muon decays, generated using FLUKA~\cite{FLUKA}. The
simulation includes all detector effects except for the presence of
impurities, whose effects are evaluated from data and included in
average in the final MC distribution~\cite{STOPMU}. The measured and
simulated energy spectra are compared in
Figure~\ref{fig:eletrackfit}. They are found to be in good agreement
($\chi^2/ndf = 14.0/20$).

We measure $\rho$ while constraining $\eta$ within its experimentally
allowed interval ($-0.020 < \eta < 0.006$)~\cite{ETA}. The spectrum
for an arbitrary pair of values, $\rho$ and $\eta$, is built by
weighting the MC events with a factor
$\frac{\frac{dP}{dx}(x_\text{MC};\rho,\eta)}{\frac{dP}{dx}(x_\text{MC};\rhoSM,\etaSM)}$,
where $x_\text{MC}=E_\text{MC}/\Emax$, and $E_\text{MC}$ is the
generated energy. We extract the value of $\rho$ as that for which the
best fit between the simulated and measured energy spectra is
obtained, which yields $\rho = 0.72 \pm 0.06$, where the error is of
statistical origin and includes the correlation with
$\eta$. Figure~\ref{fig:eletrackfit} shows the spectrum for the best
fit of $\rho$.

There are two types of source of systematic uncertainties that can
affect this measurement, namely: an underestimation of the energy
resolution and a systematic shift in the global energy scale. With the
help of the MC sample, the contributions to the total systematic error
are estimated in $\pm 0.01$ and $\pm 0.08$, respectively.  Therefore,
the total error is dominated at this level by the systematics due to
the uncertainty on the energy scale, which stresses the importance of
a fine calibration of the detector. The final result is:
\begin{equation}
\rho = 0.72\pm 0.06 \text{ (stat.)} \pm 0.08 \text{ (syst.)}
\end{equation}
compatible within the error bounds with the V-A value.

The Michel parameter $\rho$ from $\mu$ decays has been measured in the
the late 60's by several authors~\cite{RHOMU} with a precision of
about 0.4$\%$.  Such results have been obtained using dedicated
experiments involving data samples of typically several hundred
thousand events. More recently, the data from electron-positron
colliders have been used to measure the Michel parameters of the
purely leptonic $\tau$ decay~\cite{RHOTAU}. These measurements have an
accuracy of about $5\%$ and are based on the analysis of samples
typically amounting up to several ten thousand events. Our new result
is not competitive with those obtained from $\mu$ decay, and barely
with those obtained from $\tau$ decay. However, it must be remarked
that this result has been obtained using 1858 muon decay events with a
non optimized experiment. This result stresses the capabilities of the
ICARUS technology to produce robust physics results.

\section{Conclusions}

We have performed the first physics measurement with the ICARUS LAr
TPC detection technique, the Michel $\rho$ parameter, from the
detailed study of muon decay spectrum using the stopping muon event
sample from the ICARUS T600 detector test run. We have obtained $\rho
= 0.72\pm 0.06 \text{ (stat.)} \pm 0.08 \text{ (syst.)}$, in agreement
with the SM value. This measurement involves the exploitation of both
the spatial and calorimetric reconstruction capabilities of the
detector. Therefore, the obtained result constitutes a proof of the
matureness of the detection technique to produce high quality physics
results, in particular for neutrino physics when it will be installed
in the Gran Sasso underground laboratory.


\end{document}